\documentclass[trackchanges]{aastex631}
\usepackage{CJK}

\shorttitle{AASTeX v6.31 Sample article}
\shortauthors{Hu et al.}

\graphicspath{{./}{figures/}}
\usepackage{soul}
\usepackage{color,xcolor}
\usepackage{amsmath}
\usepackage{boondox-cal}
\usepackage{ulem}

\begin{document}
\begin{CJK*}{UTF8}{gbsn}

\title{Components and anisotropy of 3D QFP waves during the early solar eruption}

\author[0000-0001-9828-1549]{Jialiang Hu}
\affiliation{Yunnan Observatories, Chinese Academy of Sciences, P.O. Box 110, Kunming, Yunnan 650216, People's Republic of China}
\affiliation{University of Chinese Academy of Sciences, Beijing 100049, People's Republic of China}
\affiliation{Yunnan Key Laboratory of Solar Physics and Space Science, Kunming, Yunnan 650216, People's Republic of China}

\author[0000-0002-5983-104X]{Jing Ye}
\affiliation{Yunnan Observatories, Chinese Academy of Sciences, P.O. Box 110, Kunming, Yunnan 650216, People's Republic of China}
\affiliation{University of Chinese Academy of Sciences, Beijing 100049, People's Republic of China}
\affiliation{Yunnan Key Laboratory of Solar Physics and Space Science, Kunming, Yunnan 650216, People's Republic of China}
\affiliation{Yunnan Province China-Malaysia HF-VHF Advanced Radio Astronomy Technology International Joint Laboratory, Kunming, Yunnan 650216, People’s Republic
of China}

\correspondingauthor{Jing Ye}
\email{yj@ynao.ac.cn}

\author[0000-0002-8077-094X]{Yuhao Chen}
\affiliation{Yunnan Observatories, Chinese Academy of Sciences, P.O. Box 110, Kunming, Yunnan 650216, People's Republic of China}
\affiliation{University of Chinese Academy of Sciences, Beijing 100049, People's Republic of China}
\affiliation{Yunnan Key Laboratory of Solar Physics and Space Science, Kunming, Yunnan 650216, People's Republic of China}

\author[0000-0001-9650-1536]{Zhixing Mei}
\affiliation{Yunnan Observatories, Chinese Academy of Sciences, P.O. Box 110, Kunming, Yunnan 650216, People's Republic of China}
\affiliation{University of Chinese Academy of Sciences, Beijing 100049, People's Republic of China}
\affiliation{Yunnan Key Laboratory of Solar Physics and Space Science, Kunming, Yunnan 650216, People's Republic of China}

\author[0009-0002-6808-5330]{Shanshan Xu}
\affiliation{Yunnan Observatories, Chinese Academy of Sciences, P.O. Box 110, Kunming, Yunnan 650216, People's Republic of China}
\affiliation{University of Chinese Academy of Sciences, Beijing 100049, People's Republic of China}
\affiliation{Yunnan Key Laboratory of Solar Physics and Space Science, Kunming, Yunnan 650216, People's Republic of China}

\author[0000-0002-3326-5860]{Jun Lin}
\affiliation{Yunnan Observatories, Chinese Academy of Sciences, P.O. Box 110, Kunming, Yunnan 650216, People's Republic of China}
\affiliation{University of Chinese Academy of Sciences, Beijing 100049, People's Republic of China}
\affiliation{Yunnan Key Laboratory of Solar Physics and Space Science, Kunming, Yunnan 650216, People's Republic of China}

\begin{abstract}
The propagation of disturbances in the solar atmosphere is inherently three-dimensional (3D), yet comprehensive studies on the spatial structure and dynamics of 3D wavefronts are scarce. Here we conduct high-resolution 3D numerical simulations to investigate filament eruptions, focusing particularly on the 3D structure and genesis of EUV waves. Our results demonstrate that the EUV wavefront forms a dome-like configuration subdivided into three distinct zones. The foremost zone, preceding the flux rope, consists of fast-mode shock waves that heat the adjacent plasma. Adjacent to either side of the flux rope, the second zone contains expansion waves that cool the nearby plasma. The third zone, at the juncture of the first two, exhibits minimal disturbances. This anisotropic structure of the wavefront stems from the configuration and dynamics of the flux rope, which acts as a 3D piston during eruptions—compressing the plasma ahead to generate fast-mode shocks and evacuating the plasma behind to induce expansion waves. This dynamic results in the observed anisotropic wavefront. Additionally, with synthetic EUV images from simulation data, the EUV waves are observable in Atmospheric Imaging Assembly 193 and 211 $\AA$, which are identified as the fast-mode shocks. The detection of EUV waves varies with the observational perspective: the face-on view reveals EUV waves from the lower to the higher corona, whereas an edge-on view uncovers these waves only in the higher corona.

\end{abstract}



\section{Introduction} \label{sec:intro}

Since the first report in 1998, EUV waves have become an active  research topic in the field of solar physics. These waves are observed about 100 times annually, except for periods of solar minimum \citep{2015LRSP...12....3W}.  And these large-scale disturbances can be identified in multiple wavelengths, including soft X-ray, ultraviolet, $H_{\alpha}$, He I, white light, and radio. Some observations
 showed that EUV waves often manifest as arc-shaped or circular diffuse structures with a single peak, typically centered around solar flares \citep{2019ApJ...871..232Z}. Originating approximately 100 Mm away from the epicenter of the flare, these waves can travel at velocities up to several hundred kilometers per second, covering distances averaging about 500 Mm and occasionally spreading over the entire solar surface \citep{2019ApJ...878..106H}. \cite{2010ApJ...716L..57V} used the STEREO spacecraft to observe the dome-shaped morphology of an EUV wave, during a coronal mass ejection (CME). This study indicated that the low-coronal signatures above the limb perfectly connect to the on-disk signatures of the wave and  the dome's upward expansion speed (approximately 650 km/s) surpassing its lateral expansion rate (about 280 km/s). \cite{2011ApJ...738..160M,2019ApJ...873...22S} supported these findings, showing that the radial speed of dome-shaped EUV waves was 1.3 to 2.3 times faster than their lateral counterparts.

Early studies barely addressed the periodicity of EUV waves, because only single wavefront can be clearly observed. However, \cite{2005ApJ...633L.145B} identified oscillations with a period of several hundred seconds by analyzing intensity fluctuations along the wavefront. In a few cases, multiple wavefronts were observed, but no periodicity analysis was conducted \citep{2003SoPh..212..121H,2010A&A...522A.100P}. \cite{2011ApJ...736L..13L}, using data from the Atmospheric Imaging Assembly (AIA) on the Solar Dynamics Observatory (SDO)  conducted a detailed analysis of large-scale quasi-periodic fast propagating (QFP) wave trains within an EUV wave during an eruption. They found that the initial propagation speed of these wavefronts reached up to 1400 km/s and proposed that CME expansion was probably the primary driver behind EUV wave formation. Recently, \cite{2019ApJ...873...22S} utilized high-resolution AIA data to clearly observe the quasi-periodic multi-front structure of an EUV wave. They measured a wave period of approximately 163 s and, for the first time, determined that the wavelength of the EUV wave ranged between 84 and 110 Mm. \cite{2022SoPh..297...20S} studied several events associated with large-scale QFP waves and found that single-front EUV waves and large-scale EUV wave trains (corresponds to QFP waves here ) share similar propagation features. Therefore, two key questions remain regarding QFP waves: 1) What physical processes are responsible for the formation of QFP waves?  2) What are the specific components of each wavefront, and how solar activities influence the three-dimensional structure of these wavefronts ?

The formation of large-scale EUV waves is often related to solar eruptions. Statistical studies of intense flares indicate that even in high-energy flares, EUV waves do not occur in the absence of CMEs \citep{2006ApJ...641L.153C}. These results have led to the growing consensus that EUV waves are excited by CMEs, similar evidences are found in recent observations \citep{2012SoPh..281..187P,2020ApJ...905..150Z}.  EUV waves typically form during the early acceleration phase of the CME, with greater acceleration resulting in stronger nonlinear fast magnetoacoustic waves or shocks \citep{2012SoPh..281..187P}. As a result, large-scale EUV waves during solar eruptions are generally fast-mode shocks, produced by a piston-driven shock mechanism. However, in the real solar atmosphere, the piston effect of CME is  three-dimensional. Recnetly, \cite{2024ApJ...962...42H} conducted two-dimensional numerical simulations  to investigate QFP waves during the early stage of solar eruptions. Their results showed that arc-shaped wavefronts are successively generated and propagated outward on either side of the CME, with speeds reaching several thousand kilometers per second. By tracking the origin of the wavefronts, they found that these QFP waves originated from disturbances within the flux rope. The flux rope acted as an imperfect waveguide, allowing the perturbation inside the flux rope to escape to the outside successively via its surface, invoking the observed QFP waves. But it is still in lack of a detailed description of the structure and propagation features of QFP waves in three-dimensional space.

Observational studies of EUV waves often rely on line-of-sight integration, typically rendering wavefronts as arc-shaped or circular due to the observation direction \citep{2019ApJ...871..232Z}. Although \cite{2010ApJ...716L..57V} observed for the first time the dome-shaped structure of EUV waves using STEREO, their data could not fully reveal the complete three-dimensional structure of these waves. In this work, we employ a three-dimensional eruption model (TD model) \citep{1999A&A...351..707T} to numerically simulate the complete three-dimensional structure of QFP wavefronts during the early stages of solar filament eruptions. By tracking the origin of the wavefronts, we investigate the mechanisms responsible for QFP wave formation. We also perform a quantitative analysis of the intensity distribution and components of these three-dimensional wavefronts, providing a comprehensive characterization of their properties and origins. The structure of the paper is as follows: Section \ref{part2} introduces the model and numerical techniques used in the simulations. Section \ref{part3} presents the simulation results, including synthesized extreme ultraviolet images from the simulated physical quantities. Finally, Section \ref{part4} summarizes the findings of this study.

\section{numerical simulation}\label{part2}

To model the three-dimensional coronal eruption of a solar filament, we utilized the MPI-AMRVAC code \cite{2012JCoPh.231..718K,2014ApJS..214....4P,2018ApJS..234...30X,2023A&A...673A..66K} to solve a set of magnetohydrodynamic (MHD) equations incorporating gravity and thermal conduction . These equations are:

\begin{equation}
  \partial_{t}\rho+\nabla \cdot (\rho \textbf{v})=0\label{continum},
\end{equation}

\begin{equation}
 \partial_{t}(e)+\nabla  \cdot [(e+P^{*})\textbf{v}- (\textbf{v}\cdot \textbf{B})\textbf{B}]
  = \rho \textbf{g} \cdot \textbf{v} + \nabla  \cdot ( \eta \textbf{B} \times(\nabla \times \textbf{B})   -\textbf{F}_{c}),\label{conduction}
\end{equation}

\begin{equation}
  \partial_{t}(\rho \textbf{v})+\nabla \cdot [\rho\textbf{v}\textbf{v}+P^{*}\textbf{I}-\textbf{B}\textbf{B}] =\rho \textbf{g} ,
\end{equation}

\begin{equation}
 \partial_{t}\textbf{B}  =   \nabla \times (\textbf{v} \times \textbf{B}-\eta \nabla \times \textbf{B}),
\end{equation}

\begin{equation}
 P^{*}=p+\frac{1}{2}|\textbf{B}|^{2},
\end{equation}

\begin{equation}
 p=\rho T,
\end{equation}

These equations define the dynamics of density ($\rho$), velocity ($\textbf{v}$), magnetic field ($\textbf{B}$), pressure ($p$), energy ($e$), and temperature ($T$) within the system. We adopted the classical Spitzer model to describe thermal conduction in the plasma \citep{1962pfig.book.....S}, with the term for thermal conduction in Equation \ref{conduction} defined as $ \textbf{F}{c}=-\kappa{\parallel}( \nabla T \cdot \hat{\textbf{B}})\hat{\textbf{B}}-\kappa_{\perp}( \nabla T-( \nabla T\cdot\hat{\textbf{B}})\hat{\textbf{B}})$. The coefficients $\kappa_{\parallel}$ and $\kappa_{\perp}$ represent the thermal conduction parallel and perpendicular to the magnetic field direction, respectively. The unit vector of the magnetic field is denoted by $\hat{\textbf{B}}$, gravitational acceleration by $\textbf{g}$, and the energy term $e$ is given by: $e = \rho v^{2} /2+ p/(\gamma-1)+B^{2}/2$. All variables are dimensionless, normalized to characteristic scales of length ($L_{0}=5\times10^{9}$~cm), number density ($n_{0}=10^{10}$~cm$^{-3}$), and temperature ($T_{0}=10^{6}$~K).

The gravitational stratification of the solar atmosphere and the initial magnetic field configuration in our simulations are aligned with those documented in \cite{2023ApJ...955...88Y}. Our simulation domain spans \(8L_{0} \times 8L_{0} \times 8L_{0}\), utilizing a uniform grid with a resolution of  \(2000 \times 2000 \times 2000\) cells, each providing a spatial resolution of 200 km. This high-resolution uniform grid allows for superior capture of disturbance signals while mitigating the numerical dissipation typically observed at grid refinement boundaries, which could otherwise distort wave propagation dynamics. We employed a Godunov-style finite volume method, integrated with the Harten-Lax-van Leer approximate Riemann solver \citep{1983JCoPh..49..357H}, a third-order asymmetric flux limiter for shock capturing \citep{1995JFM...282..404S}, and a three-step Runge-Kutta time integrator. The anisotropic heat flux was addressed using the SuperTimeStepping method \citep{2012MNRAS.422.2102M}. Boundary conditions included a line-tied setup at the bottom and open boundaries on the remaining sides, facilitating unimpeded plasma flow into and out of the simulation domain.

\begin{figure}[t] 
\centerline{\includegraphics[width=1\textwidth,clip=]{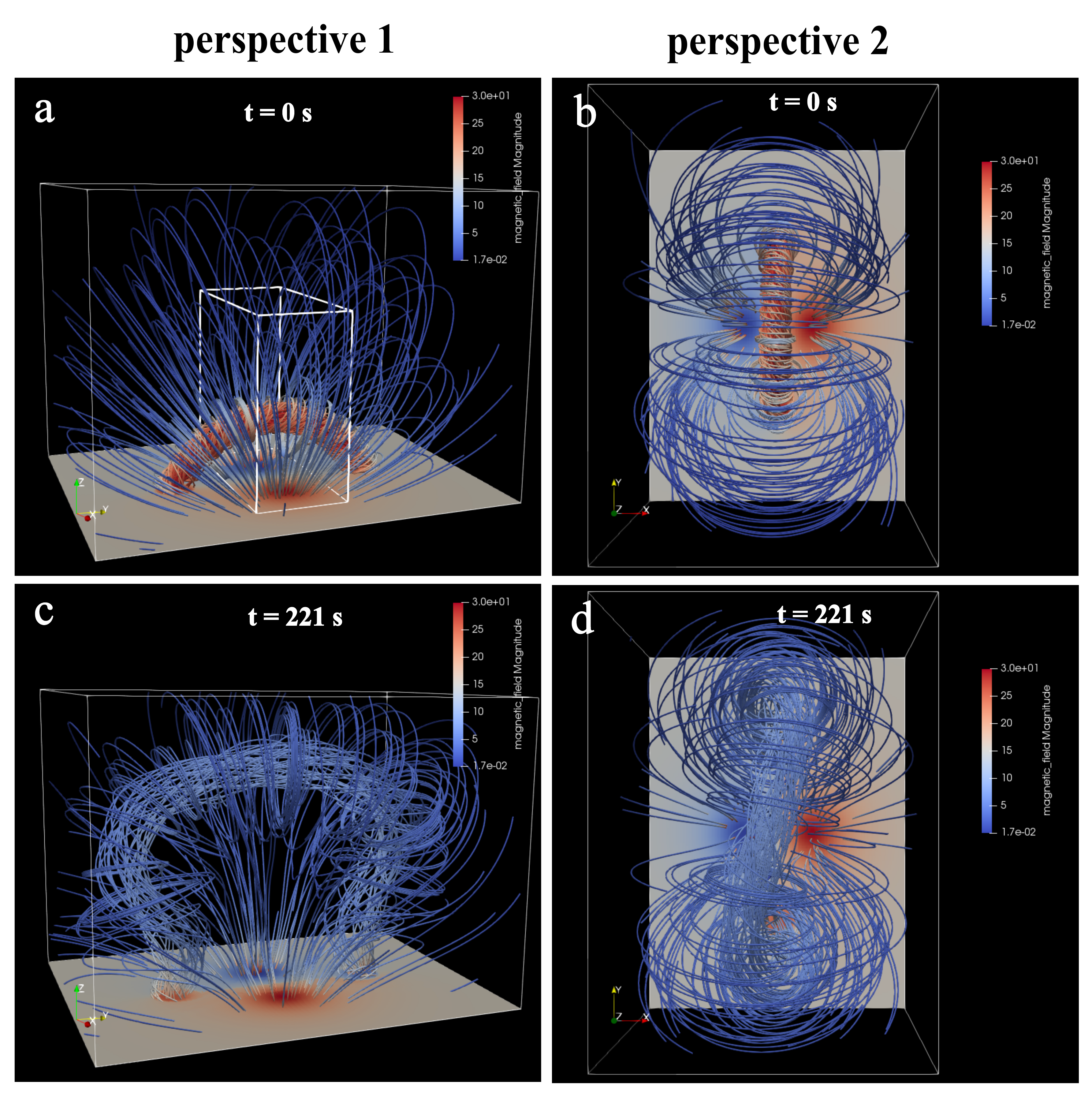}}
\caption{Overview of the magnetic field configuration from dual perspectives. This initial configuration includes three distinct components: 
$B_{I}$ , arising from the toroidal current $I$  within the flux rope; $B_{q}$, generated by two magnetic sources beneath the photosphere at depth 
$d_1$  and spaced by distance L, representing the field of an active region; and  $B_{t}$, from a dipole source at depth $d_2$ in the $x$-$y$ plane, with field strength diminishing quadratically with height. The white box in panel (a) highlights the area analyzed in Figure\ref{fig:structure}c, used to study the origin of the wave.}
\label{fig:initialmag}
\end{figure}

Figure \ref{fig:initialmag} illustrates the distribution and evolution of the three-dimensional magnetic field structure. The initial magnetic field consists of three components: $B_{I}$ , arising from the toroidal current $I$ within the flux rope;  $B_{q}$, generated by a pair of magnetic sources beneath the photosphere at depth $d_{1}$ and separated by distance L, representing an active region's magnetic field; and  $B_{t}$, produced by a dipole source at depth $d_2$  in the $x-y$ plane, with its field strength decaying quadratically with height.The initial configuration (Figures  \ref{fig:initialmag}a and  \ref{fig:initialmag}b) displays the vertical magnetic field ($B_{z}$)  at the solar surface, with field lines colored according to magnetic field strength. The initial dark red tube indicates the solar filament, anchored at two points on the solar surface along the magnetic neutral line, flanked by sunspots generating background magnetic fields that exert downward tension on the  flux rope. At t=221 s, Figures \ref{fig:initialmag}c and \ref{fig:initialmag}d illustrate changes in the magnetic field distribution. Notably, Figure \ref{fig:initialmag}c shows the flux rope experiencing a significant upward lift from its initial position, during which the internal magnetic field strength diminishes, yet new toroidal magnetic fields begin to form around the rope's axis. Concurrently, Figure \ref{fig:initialmag}d presents the flux rope’s magnetic field adopting an S-shaped configuration, a morphology consistent with findings from previous research \citep{2016cosp...41E2181Z}. During this upward displacement of the flux rope, the Quasi-Periodic Fast Propagating (QFP) wave phenomenon is generated. A comprehensive analysis of the QFP wave results will be elaborated in the subsequent section.

\section{Results}\label{part3}

This section offers an in-depth examination of the propagation characteristics and generation mechanisms of quasi-periodic fast propagating (QFP) wave trains during the initial phases of solar eruptions.

\subsection{Structure and Origin of 3D Wavefronts }\label{section1}

\begin{figure}[t] 
\centerline{\includegraphics[width=1\textwidth,clip=]{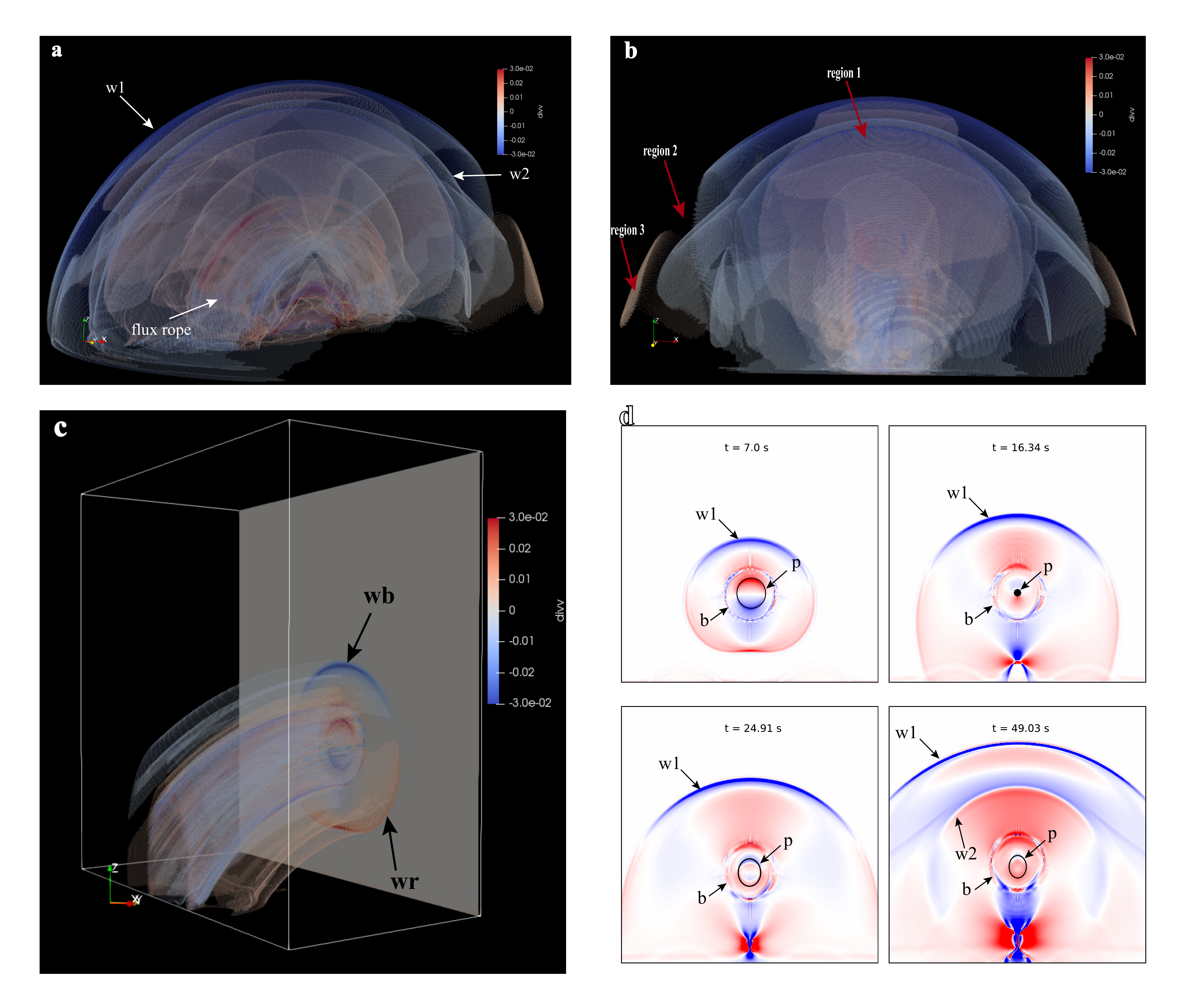}}
\caption{Analysis of the three-dimensional propagation processes and mechanisms of QFP waves. Panels (a) and (b) depict the propagation of velocity divergence from two perspectives, highlighting the first (w1) and second (w2) wavefronts, respectively. Panel (c) explores disturbance propagation in a localized area near the flux rope, tracing the origins of these wavefronts, with 'wb' and 'wr' denoting the upper and lower segments of the initial wavefront w1. Panel (d) presents a temporal evolution diagram of the background slice in Panel (c), captured at different time intervals, "b" denotes the boundary of the flux rope, "p" represents the disturbance inside the rope, and "w1" and "w2" indicate the two generated wavefronts (corresponding to the two wavefronts in Panel \ref{fig:structure}a). An accompanying video further elucidates the dynamic propagation process, offering a more intuitive visualization of the wavefront evolution.}
\label{fig:structure}
\end{figure}


Figure \ref{fig:structure} display the velocity divergence maps following the eruption, which are used to characterize the generation and propagation of three-dimensional waves. Panels a and b show the distribution of velocity divergence at t=70~s from two different perspectives. Positive divergence values indicate plasma compression, while negative values represent plasma expansion. Thus, the 3D distribution of velocity divergence reflects  plasma disturbances near the wavefronts.

Panel a and the associated video show the upward eruption of the flux rope and the resultant wave phenomena. Initially, the flux rope becomes unstable, and the Lorentz force produced by the background magnetic field acts on the edge of the rope. 
This background magnetic field, anchored in the photosphere, reconnects beneath the flux rope, reducing its confining influence and enabling the flux rope's upward acceleration. Throughout this ascent, multiple wavefront structures are generated, with the first and second labeled as "w1" and "w2," respectively. Computational analysis reveals that these wavefronts propagate at a speed of 1282 km/s, aligning with typical EUV wave speeds (200$\sim$1500 km/s)  \citep{2014SoPh..289.4563M} and closely matching the 1240 km/s speed observed in large-scale QFP waves \citep{2011ApJ...736L..13L}. Furthermore, the simulation reveals that the second and third wavefronts emerge from the flux rope surface at t=36 s and t=72 s, respectively, yielding a period of approximately 36 s for the QFP waves, which is similar to the observations by \cite{2019ApJ...871L...2M} (with a period of 45 s). It is noteworthy that the shape of the three-dimensional wavefront is not a spherical structure centered at the eruption site, but rather an ellipsoidal shape that spreads outward around the flux rope, tightly associated with the shape of the erupting filament.  The video associated with Figure \ref{fig:structure}a shows that the 3D wavefronts expands outward in both radial and lateral directions in a dome-like shape, with the radial part propagating faster than the lateral part. This phenomenon is consistent with observations in previous events \citep{2010ApJ...716L..57V,2019ApJ...873...22S}. This indicates that the upward motion of the  flux rope plays a significant role in shaping the three-dimensional structure of the wavefronts. As the propagation process continues, the entire wave structure becomes more complex.

Panel b and its related video illustrates the propagation of waves observed along the axial direction of the flux rope, revealing the propagation of QFP waves at the flanks of the flux rope. An intriguing and noteworthy phenomenon observed in Panel b is that a complete wavefront in the solar atmosphere can be divided into three distinct regions based on the velocity divergence: (1) region with velocity divergence less than 0 (marked as the blue area "region 1" in Figure \ref{fig:structure}b): This is the compressive region, primarily located ahead of the erupting flux rope and constituting the major portion of the entire wavefront.
(2) region with velocity divergence greater than 0 (marked as the red area "region 3" in Figure \ref{fig:structure}b): This is the expansive region, situated at the flanks of the flux rope and covering a relatively smaller area within the entire wavefront.
(3) region with velocity divergence equal to 0 (marked as "region 2" in Figure \ref{fig:structure}b): This region, located at the boundary between the blue and red areas, exhibits zero velocity divergence and occupies the smallest portion of the wavefront. These regions suggest that the three-dimensional wavefront comprises different components at different heights and directions. The properties of these components will be explored in detail in Section \ref{section2}.

In Figures \ref{fig:structure}a and \ref{fig:structure}b, we observe the formation of multiple QFP wavefronts during the initial stages of an eruption. To investigate the generation mechanism of these QFP waves, we refer to the study by \cite{2024ApJ...962...42H} and focus on a localized region near the flux rope (highlighted by the white box in Figure \ref{fig:initialmag}a) to examine the propagation of disturbances. Figure\ref{fig:structure}c and the related movie display the formation and propagation of these QFP waves within this local region. The movie clearly reveal the movement of disturbances inside the flux rope and its interaction with surrounding  structures. Figure \ref{fig:structure}d captures several critical moments in the evolution of these waves. 

From t = 7.0 to 16.34 seconds, the circular disturbance "p" propagates from the boundary of the flux rope towards its center, gradually contracting to a point,  and simultaneously initiating the outward propagation of the first wavefront ,"w1", into the corona. Between t = 16.34 and 24.91 seconds, the disturbance "p"  reverses direction, moving from the center back towards the boundary. Upon  reaching the boundary,  the disturbance "p" bifurcates: part transmits outward while the remainder reflects back towards the center. At t = 49.03 seconds, the second wavefront (labeled as "w2") is observed propagating outwards, while the reflected disturbance "p" again moves inward. This cyclical disturbance pattern offers a compelling  explanation for the event observed in active region 11105 on September 8-9, 2010 \textcolor{red}{\citep{2012ApJ...753...52L}}, where sequential large-scale wavefronts were detected on the lateral CME. In their time-distance plots from azimuthal cuts (see Figure 5e in Liu et al. \cite{2011ApJ...736L..13L}), multiple wavefronts emerging from the prominence surface at nearly the same speed were visible, indicating that the flux rope acts as an imperfect waveguide for generating large-scale QFP waves. This result aligns with the mechanism described in two-dimensional models, and a detailed analysis can be found in \cite{2024ApJ...962...42H}. However, with the same excitation, the production of waves can be significantly different in 3D than 2D.

\subsection{Quantitative analysis of anisotropy of 3D wavefront}\label{section2}
In the previous section, we identified an inhomogeneous distribution of velocity divergence across the wavefront, indicating that various heights might host distinct wave components. This section delves into a detailed examination of these components,  focusing on their correspondence to different regions of wavefront and exploring the underlying reasons for their spatial distribution.
 
 To quantitatively analyze the intensity and components of the wavefronts within the three-dimensional spatial domain, we select two mutually perpendicular cut planes: Plane 1 (yz plane at x= 0), passing through the axis of the flux rope, and Plane 2 (xz plane at y=0), perpendicular to the axis. Notably, Plane 1 intersects only region 1, as depicted in Figure \ref{fig:structure}b, while Plane 2 crosses all three designated regions. Our analysis will focus on the properties of the waves located on these two planes.
  
\begin{figure}[t] 
\centerline{\includegraphics[width=1\textwidth,clip=]{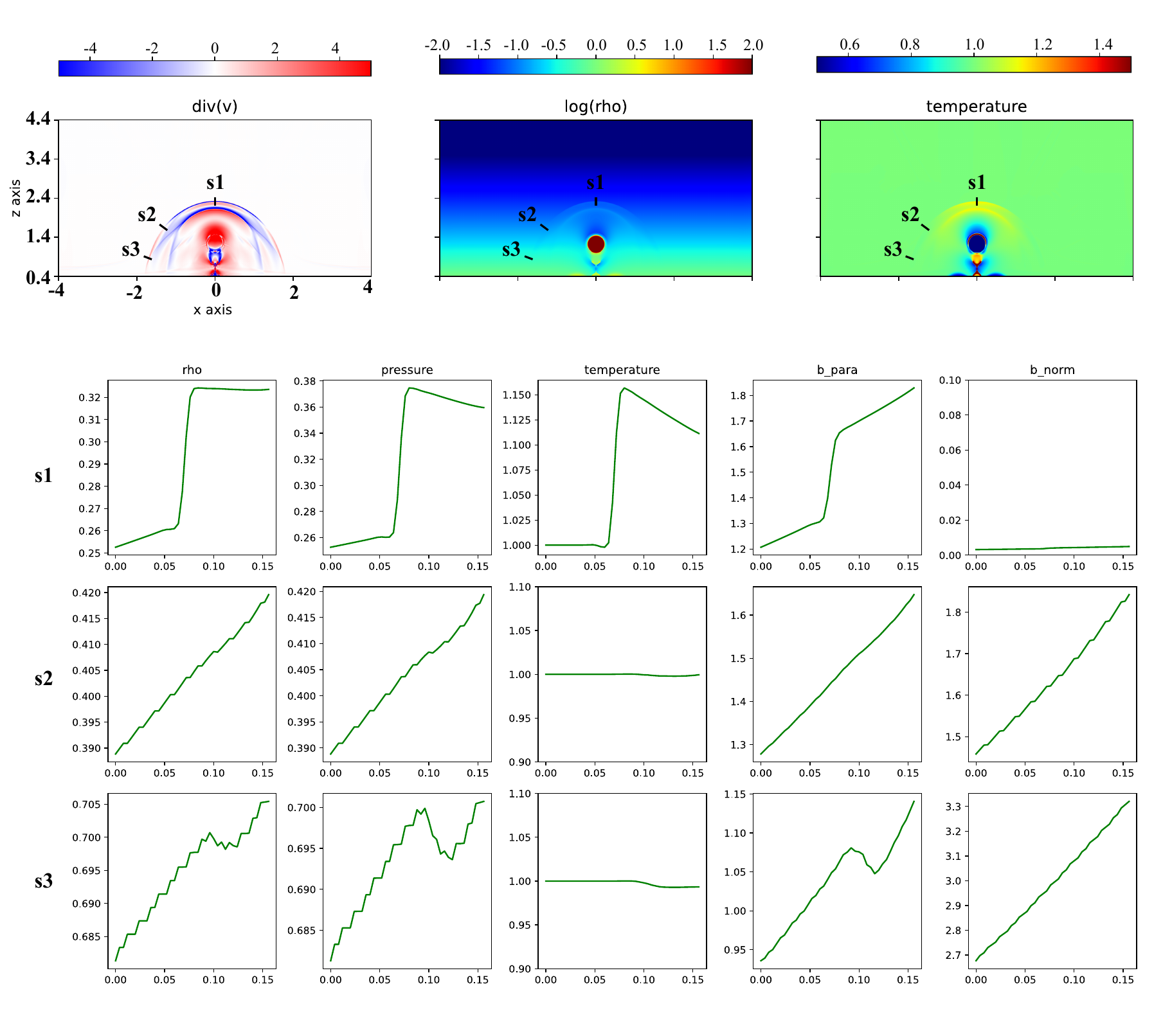}}
\caption{Analysis of the properties of QFP waves on plane 2 (the xz-plane at y=0). The top row shows the distributions of velocity divergence, logarithmic density, and temperature. We selected three slices—s1, s2, and s3—on the first wavefront of the QFP wave to quantitatively analyze the variation of physical quantities at different positions across the wavefront. The next three rows present the distributions of density, pressure, temperature, the magnetic field component parallel to the wavefront, and the magnetic field component perpendicular to the wavefront along slices s1, s2, and s3, respectively.}
\label{fig:xzplane}
\end{figure}

\begin{figure}[t] 
\centerline{\includegraphics[width=1\textwidth,clip=]{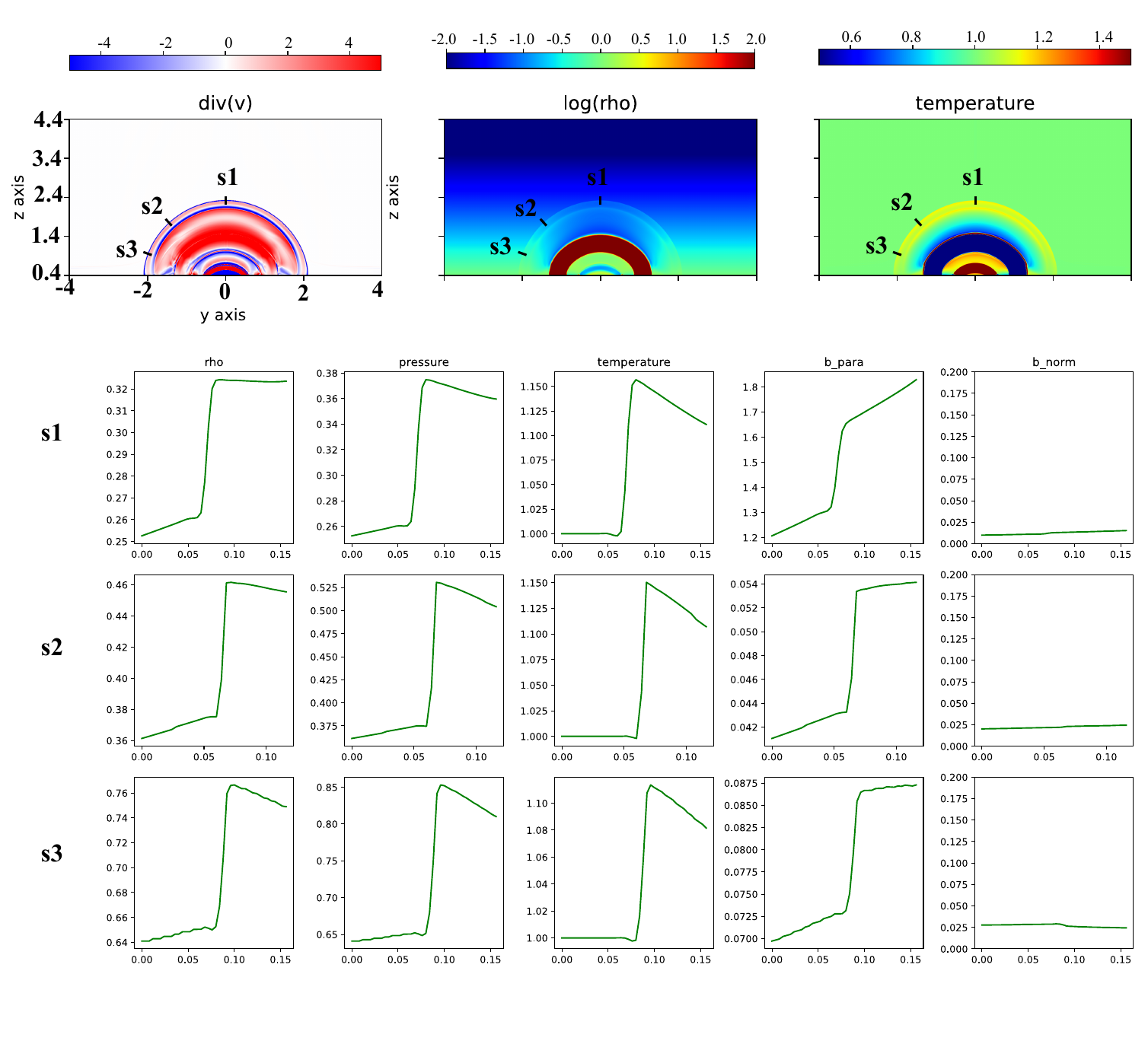}}
\caption{The yz-plane at x=0 (plane 1). The physical meanings of the quantities shown in this figure are the same as those in Figure \ref{fig:xzplane}.}
\label{fig:yzplane}
\end{figure}

 The first rows of Figures \ref{fig:xzplane} and \ref{fig:yzplane} illustrate the distributions of velocity divergence, logarithmic density, and temperature across the wavefronts. Velocity divergence maps effectively reveal the spatial distribution of the entire wavefront. In Figure \ref{fig:xzplane}, the density and temperature profiles distinctly capture only the wavefront structure above the flux rope. Conversely, 
 Figure \ref{fig:yzplane} provides a comprehensive view of the entire wavefront. The discrepancy occurs because significant various in density and temperature are primarily manifest within the wavefront located in region 1.
 
In Figure \ref{fig:xzplane} and Figure \ref{fig:yzplane}, the wavefronts observed within region 1 consistently exhibit characteristics of shock waves, with significant increase in temperature, pressure, and density. For instance, in the second row of Figure \ref{fig:xzplane}, slice s1 displays density, pressure, and temperature compression ratios of 1.23, 1.5, and 1.15, respectively. Similarly, in Figure \ref{fig:yzplane}, the compression ratios across slices s1,s2,s3 show slight variations, with density ratios around 1.2 and temperature ratios approximately 1.1. Pressure ratios, however, vary by location.
Further analysis reveals that the magnetic component parallel to the wavefront significantly increases after the passage of the wavefront, while the perpendicular component remains almost unchanged, reinforcing that these wave are fast-mode shock waves. The simulation's shock wave parameters align closely with those observed in a fast-mode shock wave event on February 15, 2011 \citep{2011ApJ...738..167S}, where density and temperature increases were approximately 10\%
and 7\%, respectively - comparable to our findings of about 15\% and 10\%.  Additionally, the intensity of these shock waves in extreme ultraviolet (EUV) is often quantified using the Mach number, typically estimated from wave speed, density compression ratios, and type II radio burst band splitting. Prior studies indicate that fast-speed EUV waves have initial Mach numbers greater than 1.15 \cite{2019ApJ...873...22S}, while moderate-speed waves fall below 1.04 \citep{2011ApJ...727L..43K}. Our simulation suggests a Mach number of 1.28, consistent with the characteristics of fast-speed EUV waves, emphasizing the conversion of wave energy into thermal energy, thereby heating the surrounding coronal plasma.

In Figure \ref{fig:xzplane}, slice s3 is situated within region 3 of Figure \ref{fig:structure}b, where density and temperature experience slight decreases after the passage of the wavefront, indicating characteristics of an expansion wave (or rarefaction wave). Despite a modest temperature reduction of only 0.3\%, this decreases still contributes to cooling the surrounding plasma, contrasting with the heating effect typically associated with compression waves. Moreover, we found that expansion waves modify the structure of the magnetic field. Following the passage of an expansion wave, there is a decrease in the magnetic field component parallel to the wavefront, whereas the perpendicular component remains largely unaffected. 

Slice s2 in Figure \ref{fig:xzplane}, located in region2 of Figure \ref{fig:structure}b, corresponds to the interface between the expansion and compression waves, where the velocity divergence is zero. Seen from the second-to-last row of Figure \ref{fig:xzplane},  there is no change in density, temperature, or magnetic field before and after the passage of the wave front. The stability in physical quantities is primarily due to  Slice s2's unique position at the boundary between opposing wave types. Here, the compression and expansion effects counteract each other, effectively neutralizing disturbances.

In summary, the wavefronts in region 1 primarily consist of fast-mode shock waves, concentrated ahead of the flux rope's motion. Conversely, the wavefronts in region 3 are characterized as expansion waves that propagate at the flanks of the flux rope. The distinct wave components  in different directions across the  wavefront -highlighting its anisotropic nature- will be further analyzed in the following section.

\subsection{The Cause of anisotropy at the 3D wavefront}\label{section3}

\begin{figure}[t] 
\centerline{\includegraphics[width=1\textwidth,clip=]{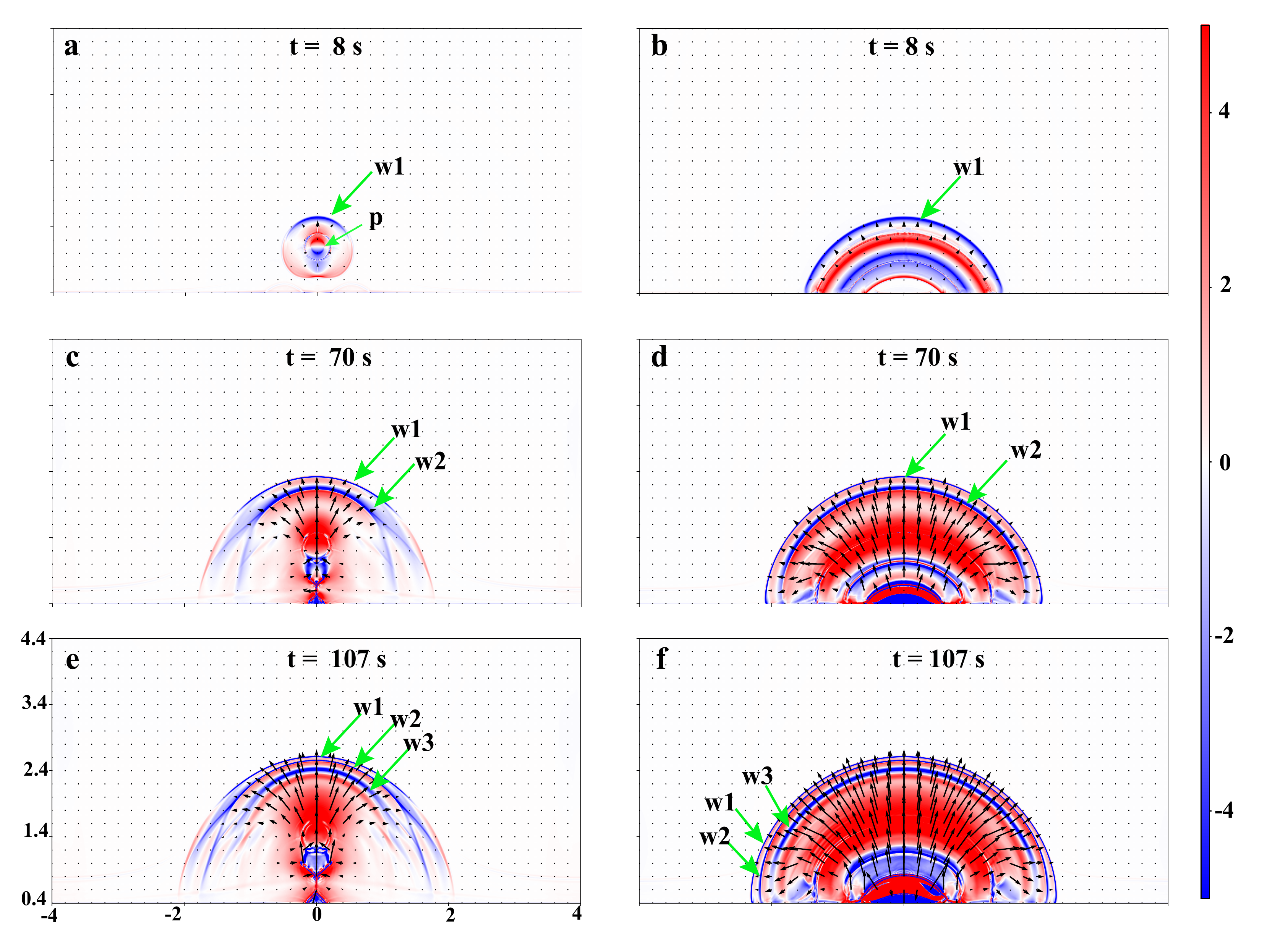}}
\caption{Velocity divergence and velocity distribution on  plane1 and plane2.}
\label{fig:velocity_distribution}
\end{figure}

In our previous analysis, we delineated three distinct wave components within the wavefront: fast-mode shocks in region 1, expansion waves in region 3,  and a stable interface region with no significant perturbations. This section delves deeper into the origins of this anisotropic three-dimensional wave composition, aiming to uncover the mechanisms driving these varied wave phenomena across different regions.

The formation of large-scale EUV waves is generally attributed to CMEs via a piston-driven shock mechanism that typically activates during the initial acceleration during the initial acceleration phase of the CMEs. Acceleration plays a critical role in shaping these waves, with greater accelerations fostering stronger nonlinear fast magnetoacoustic waves or shocks \citep{2010ApJ...718..266M}. Unlike the simplified one-dimensional piston models often discussed, the piston effect in the solar atmosphere manifests in three dimensions. As a prominence accelerates upward, it not only compresses the plasma above but also initiates an evacuation effect beneath it, creating a dual dynamic. This interplay between compression above and evacuation below is characteristic of the three-dimensional piston effect, which naturally should lead to the generation of expansion waves alongside compression waves. 
 
To elucidate the evolution and characteristics of different wave components at the wavefront, we examine the velocity divergence and flow field motions at different times, as depicted in Figure \ref{fig:velocity_distribution}. In Figure \ref{fig:velocity_distribution}a, at t = 8 s,  we observe the emergence of a  compressed fast-mode shock wave (corresponding to the blue part of "w1") appears ahead of the flux rope. This shock wave formation is  primarily driven by the upward acceleration of the flux rope. Simultaneously, behind the flux rope, an expansion wave (corresponding to the red part of "w1" in Figure \ref{fig:velocity_distribution}a) forms as a result of the evacuation effect created by the flux rope's motion. Additionally,  the internal dynamics of the flux rope,  labeled as disturbance 'p', respond to this upward acceleration.   The upper part of  "p" expands (corresponding to the red part of "p" in Figure \ref{fig:velocity_distribution}a), while the lower part compresses (corresponding to the blue part of "p" in Figure \ref{fig:velocity_distribution}a). This is determined by the direction of disturbance propagation and the direction of flux rope acceleration. For the upper part of "p", the direction of disturbance propagation is opposite to the direction of flux rope motion, hence the expansion. Conversely, the direction of the lower part of the disturbance is the same as the direction of flux rope motion, resulting in compression.  As the wavefront, "w1", progresses to t = 70 s, shown in  Figure \ref{fig:velocity_distribution}c, it evolves to include both the forward fast-mode shock waves and the flank's expansion waves.
  
The velocity field distribution elucidates the significant role of the flux rope's upward motion in shaping fast-mode shock wave ahead of it. The flow is primarily concentrated in two distinct regions. Firstly, near the current sheet located directly beneath the flux rope, weaker inflow motions can be observed (as shown in Figures \ref{fig:velocity_distribution}c and e ). These inflows facilitate magnetic reconnection at the current sheet,  which reduces the confining forces exerted by the background field on the flux rope, thereby enabling its upward motion. Secondly, the region above the flux rope displays a pattern where the velocity arrows spread outward, indicating the diffusion of the flux rope's upward motion to both flanks  (seen from  Figures \ref{fig:velocity_distribution}a ,c to e). The velocity disturbances from the flux rope's movement extend until they encounter the leading edge of the fast-mode shock wave. This clear propagation path illustrates the spatial correlation between the flux rope's ascent and the resultant shock wave development, underlining the flux rope's role in shaping the large-scale dynamics of the surrounding plasma.

Figures \ref{fig:velocity_distribution}b, d, and f provides a good view of the flux rope's structure and dynamics. As the flux rope ascends from t = 8 to 70s, the diffuse wavefront ("w1" in Figures \ref{fig:velocity_distribution}b) gradually becomes sharper ("w1" in Figures \ref{fig:velocity_distribution}d). This sharpening exemplifies the steepening process characteristic of EUV waves, not only observed in high-resolution AIA observations \citep{2010ApJ...723L..53L} but also in simulations \cite{2009ApJ...700.1716W}. As the flux rope ascends from t = 70 to 107s, the distance between the upper part of  "w1" and  "w2" in front of the flux rope shortens, meaning that the latter wavefront catches up with the former  (see Figures \ref{fig:velocity_distribution}d and f).  This is a direct consequence of the flux rope’s upward motion. Given that the motion of the flux rope is predominantly radial, it exerts a lesser impact on lateral wave propagation. Consequently, Figure \ref{fig:velocity_distribution}f illustrates a notable disparity in wave dynamics: the distance between the lateral parts of these wavefronts is significantly greater than the upper parts. This behavior indicates the directional impact of the flux rope's motion on the propagation of waves.

In summary, the distribution of different components across various regions on a three-dimensional wavefront is jointly determined by the structure and motion of the flux rope. During the initial stages of the flux rope eruption, the rope acts like a three-dimensional piston. On one hand, it compresses the disturbances ahead of the piston to generate fast-mode shocks, and on the other hand, it evacuates the plasma behind the piston to produce rarefaction (or expansion) waves. As the entire wavefront propagates, distinct wave patterns emerge in different regions, as shown in Figure \ref{fig:structure}b. Since the upward motion direction of the flux rope aligns with the propagation direction of the fast-mode shock in Region 1, the size of the fast-mode shock region is determined by the motion direction of the flux rope. In contrast, the rarefaction waves in the later stages of propagate along the solar surface, nearly perpendicular to the downward evacuation direction of the flux rope. Consequently, their propagation intensity weakens significantly over time.

\subsection{Synthetic EUV images along different light of sight }\label{section4}

\begin{figure}[t] 
\centerline{\includegraphics[width=1\textwidth,clip=]{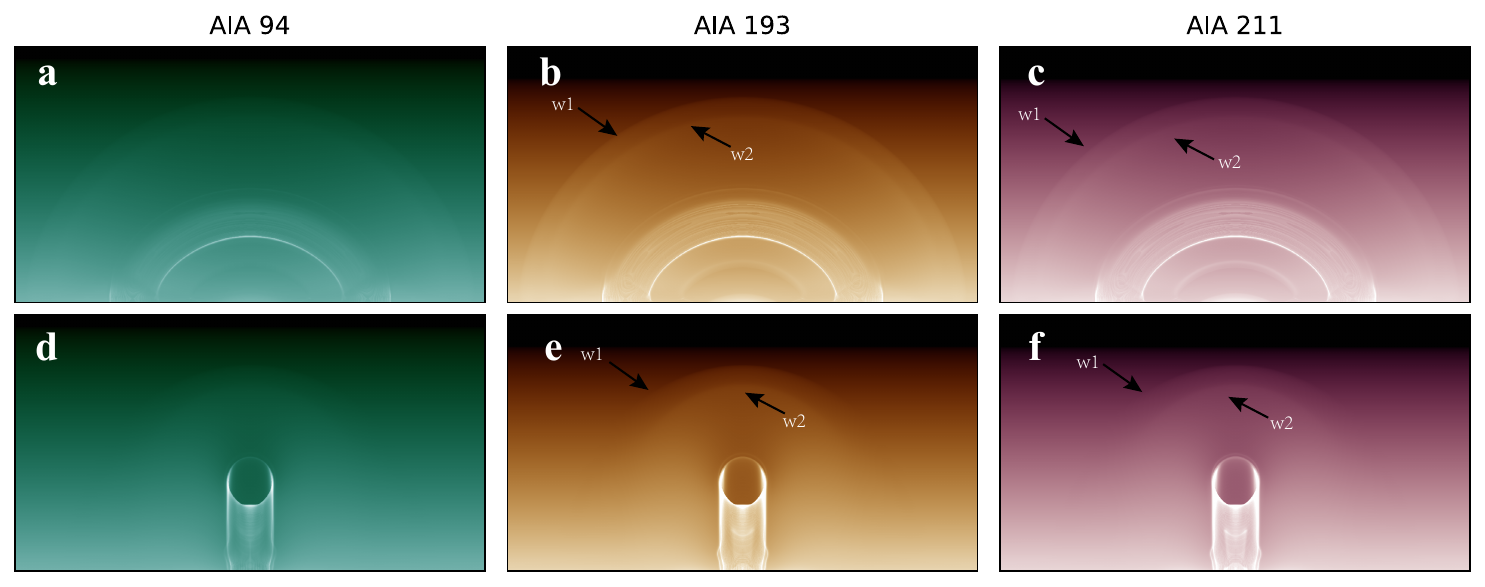}}
\caption{Synthesized images from the AIA 94 $\AA$ 193 $\AA$, and 211 $\AA$ bands. The first and second rows depict different viewing angles, respectively, with "w1" and "w2" indicating the two wavefronts of QFP waves.}
\label{fig:syn}
\end{figure}



In this study, we employ the FoMo 3.4 program \citep{2016FrASS...3....4V} to synthesize EUV emissions. The synthetic images focus on the 193 , 94  and 211 $\AA$ bands, and they include two line-of-sight directions, as shown in Figure \ref{fig:syn}. First, the synthesized results of the wavefront are similar to actual observational events. The response to coronal disturbances is more pronounced in the 193 $\AA$ and 211 $\AA$ bands, while it is weaker in the 94 $\AA$ band. This phenomenon indicates that the plasma heating induced by EUV waves primarily occurs within the temperature range of 0.8 to 1.6 MK \citep{2014SoPh..289.3233L}.

Second, our findings reveal that observing the same EUV wave from different perspectives yields distinct outcomes. When observing the wave phenomenon along a direction parallel to the axial direction of the flux rope (as shown in Figures \ref{fig:syn}e and  \ref{fig:syn}f), the wave is predominantly concentrated above the flux rope, appearing as elongated arcs in the low coronal region (marked as "w1" and "w2"). There is no response at the flanks. Conversely, when  observing the wave perpendicular to the flux rope's axial direction, the entire semicircle wavefront, extending from the chromosphere to the  corona, becomes visible. 

Previous research by \cite{2012ApJ...752L..23S}using SDO/AIA data demonstrates the wave consistency across the photosphere, chromosphere and low corona. Generally, observational results indicate that for the same event, wave signals at different atmospheric levels are largely similar, reflecting the activity characteristics of the same EUV wave at various heights \citep{2015LRSP...12....3W}. However, not all EUV waves can produce significant responses in lower atmosphere, and our synthesized images show that the view angle (or LOS, line of sight) significantly influence the EUV wave responses to different heights.  Some angles allow detection of EUV signals from the lower solar atmosphere to the upper, as seen in in Figures \ref{fig:syn}b and  \ref{fig:syn}c. However, some angles only capture EUV wave in the upper atmosphere, as illustrated in Figures \ref{fig:syn}e and  \ref{fig:syn}f. This phenomenon occurs because the region that can produce signals responses to  EUV waves correspond to the fast-mode shock region of the entire wavefront, while there is no response to the rarefaction wave. The area of the fast-mode shock is limited and does not cover the entire wavefront, leading to different heights responding to EUV  depending on the LOS.

Furthermore, upon examining Figures \ref{fig:syn}b and \ref{fig:syn}c, we notice a significant reduction in the intensity of the second wavefront "w2" compared to the first wavefront "w1". This observation concurs with the research findings presented in \cite{2011ApJ...736L..13L} (specifically, Figure 8b therein), which clearly demonstrates that the brightness of the second wavefront emerging from the filament's side is considerably dimmer than that of the first wavefront.

\section{Summary And Discussions} \label{part4}

In observations, EUV waves often appear as two-dimensional arc-like structures, as demonstrated by the numerical studies of \cite{2009ApJ...700.1716W} and \cite{2012SCPMA..55.1316M}. However, in reality, wave propagation in the solar atmosphere occurs in three dimensions. The current understanding of the structure and components of three-dimensional wavefronts remains unclear. Based on the analytical model by \cite{1999A&A...351..707T}, we conducted numerical experiments to study the three-dimensional structural characteristics of QFP waves and their generation mechanism during the early stage of a solar filament eruption. The main conclusions are as follows:

1. \textbf{Three-Dimensional Wavefront Characteristics}: We provide the first detailed depiction of the three-dimensional wavefronts of EUV waves, which form a dome-shaped region above the flux rope. This  structure is composed of three distinct regions as illustrated in  Figure \ref{fig:structure}b: the fast-mode shock region directly in front of the flux rope, the expansion wave region at the flanks of the flux rope, and a transitional boundary between them. The fast-mode shock effectively converts wave energy into thermal energy, heating the plasma, whereas the expansion wave contributes to localized cooling.

2. \textbf{Dynamics of the Flux Rope as a 3D Piston}: The flux rope acts as a three-dimensional piston. Its upward acceleration during the eruption compresses the plasma ahead, generating fast-mode shocks, while simultaneously inducing expansion waves through rarefaction behind it. This dynamic results in the formation of three defined regions on the wavefront of the dome-shaped structure as the wavefront propagates.

3. \textbf{Propagation and Origin of Wavefronts}: Our analysis shows that fast-mode shocks propagate at velocities up to 1240 km/s and that the QFP waves, with a periodicity of approximately 36 s, originate from disturbances within the flux rope. This flux rope acts as an imperfect waveguide, effectively confines the disturbance inside. When the disturbance reach the boundary of flux rope, it partially penetrate the surrounding corona, manifesting as observable QPF waves.

4. \textbf{Observational Implications of Synthetic Imaging}: Synthetic imaging analysis indicates that the EUV waves are observable in AIA 193 and 211 $\AA$, which are identified as the fast-mode shocks. The visibility of the EUV wave varies with the observation angle -- some perspectives reveal a complete wavefront extending from lower to higher corona, while some only detect signals from higher corona. This variability underscores the importance of the three-dimensional shock distribution and LOS.

The filament here is represented by a ring structure that contains a  magnetic flux rope. As for those filaments formed by irregular shapes, the distribution characteristics of their fast-mode shock waves remain unclear at present. However, it is certain that this distribution is related with  the shape and motion of the filament. In future, we will aim to model  three-dimensional irregular wave structures using advanced techniques like flux rope insertion method \citep{2004ApJ...612..519V,2017SPD....4840606T}, addressing potential variations in the fast-mode shock regions for irregular filaments.

\vspace{2em}
\noindent
This work was supported by Strategic Priority Research Program of the Chinese Academy of Sciences No.XDB0560000, NSFC grants 11933009, 12273107, U2031141,and 12073073; grants associated with the Yunling Scholar Project of the Yunnan Province, the Yunnan Province Scientist Workshop of Solar Physics, the Yunnan Key Laboratory of Solar Physics and Space Exploration of Code 202205AG070009, the Special Project for the construction of science and technology innovation centers faced to South Asia and Southeast Asia-Yunnan International Joint Innovation Platform: " China-Malaysia HF-VHF Advanced Radio Astronomy Technology International Joint Laboratory of Yunnan" (202303AP140003). J.Y. and Z.M. also acknowledge the support by grants associated with the Yunnan Revitalization Talent Support Program and the Foundation of the Chinese Academy of Sciences (Light of West China Program). The numerical computation in this paper was carried out on the computing facilities of the Computational Solar Physics Laboratory of Yunnan Observatories (CosPLYO).

\bibliography{bibtex}{}
\bibliographystyle{aasjournal}

\end{CJK*}
\end{document}